\documentclass[a4paper,11pt]{article}
\pdfoutput=1 % if your are submitting a pdflatex (i.e. if you have
\usepackage{jcappub} % for details on the use of the package, please
\usepackage[T1]{fontenc} % if needed

\title{\boldmath On the Cosmological Constant-Graviton Mass correspondence}

\author[a,1]{Oem Trivedi,\note{Corresponding author.}}
\author[b,2]{Abraham Loeb}

\affiliation[a]{International Centre for Space and Cosmology, Ahmedabad University, \\ Ahmedabad 380009, India }
\affiliation[b]{Astronomy Department, Harvard University, 60 Garden  St., Cambridge, \\ MA 02138, USA}

\emailAdd{oem.t@ahduni.edu.in}
\emailAdd{aloeb@cfa.harvard.edu}

\abstract{Relations between the graviton mass and the cosmological constant $\Lambda$ have led to some interesting implications. We show that in any approach which leads to a direct correlation between the graviton mass and $\Lambda$, either through direct substitution of gravitational coupling in dispersion relations or through the linearization of Einstein equations with massive spin-2 fields, the Compton wavelength of the graviton lies in the superhorizon scale. As a result any gravitational approaches where the graviton mass is related directly to the cosmological constant are of no observational significance.}

\begin{document}
\maketitle
\flushbottom

\section{Introduction} \label{sec:intro}
The notion of a nonzero graviton mass has garnered substantial attention in recent years, both as a fundamental question in high energy physics and for its potential implications in cosmology and astrophysics. In standard formulations of general relativity(GR), gravitons are treated as massless spin-2 gauge bosons, consistent with GR's invariance under diffeomorphisms and the consequent long-range nature of gravitational forces. However, alternative approaches which include modifications of GR and extensions such as massive gravity theories, allow for a small but non-zero mass for gravitons, $m_{g}$(see \cite{de2017graviton} for an extended review). These non-GR frameworks propose mechanisms by which graviton mass could emerge, which could even introduce modifications to the inverse-square law at cosmological scales and influence the large-scale structure of spacetime. 
\\
\\
Significant interest has hence been laid towards finding appropriate bounds on the graviton masses, based on the LIGO collaboration \cite{abbott2021tests} and other constraints \cite{Rana:2018vxn,mandal2024origin,particle2022review}. Recently, a stringent bound on the graviton mass was derived \cite{loeb2024new} using measurements of the CMB dipole \cite{aghanim2014planck}. \\
Attempts were made to relate the cosmological constant with the graviton mass. Such a relation was firstly found by using a linearization approach to the Einstein equations in a de Sitter background \cite{novello2003mass}. Here we show that one can arrive at such a relation using much simpler considerations as well. We then discuss the observational consequences of the results.
\section{Cosmological Constant-Graviton Mass Correspondence}
An initial approach towards a direct correspondence between the graviton mass,$m_{g}$ and the cosmological constant,$\Lambda$, was considered in \cite{novello2003mass}, with further work in \cite{liao2004gravitational,tajmar2006note,espriu2021effect}. The derived relation in this case is, \begin{equation} \label{cc}
    \Lambda = \frac{3 m_{g}^2 c^2}{2 \hbar^2 }
\end{equation}
This follows naturally from linearizing Einstein’s field equation for gravity including a cosmological constant as well as from the equations of motion for a massive spin-2 field propagating in a de-Sitter background. It was also shown that the field has only two degrees of freedom when the mass is of the form specified by \eqref{cc} \footnote{The reader may note that while it is true that
generically massive gravitons propagate 5 degrees of freedom and it is
also true that for\eqref{cc} it propagates fewer degrees of freedom,
but contrary to \cite{novello2003mass}, there could be 4 degrees of freedom here instead of 2 \cite{Deser:2001wx}.} and five degrees of freedom otherwise. This result again comes from essentially Einstein gravity and remains true also if the background field equations are Einstein equations with $\Lambda$ \cite{novello2005mass}. The claim is that in order for one to not induce extra degrees of freedom, at least in a dS background, the graviton mass must be of the form \eqref{cc}. 
\\
\\
One can consider an alternate approach towards reaching at a relation similar to \eqref{cc} as well. The plasma dispersion relation for photons introduces an effective photon mass. Specifically, in a plasma the dispersion relation is modified to, \begin{equation}
    \omega^2 = k^2 c^2 + \omega_p^2
\end{equation} where $\omega$ is the photon frequency, k is the wave vector, and c is the speed of light. This is equivalent to a relativistic energy expression, $E^2 = p c^2 + m_\gamma^2 c^4$ and here, the effective photon mass $m_\gamma$ is given by $m_\gamma^2 = \frac{\hbar^2}{c^4} \omega_p^2$, with $\omega_p = \sqrt{\frac{4 \pi e^2 n_{e}}{m_{e}}}$ as the plasma frequency, where e is the electron charge, $n_{e}$ is the number density of electons and $m_{e}$ is the electron mass. For gravitons, the gravitational coupling can be analogously defined by replacing the electromagnetic coupling $e^2$ with $G m^2$, where $G$ is the gravitational constant. This substitution implies a graviton mass $m_{g}$ that satisfies, \begin{equation}
    m_g^2 = \frac{\hbar^2}{c^2} \times 4 \pi G \rho
\end{equation}
where $\rho$ is the mass density. Using the density parameter $\Omega$, defined as $\Omega = \frac{\rho}{3 H_0^2 / 8 \pi G}$ where $H_0$ is the Hubble constant, we can further express $m_g$ in terms of cosmological parameters which yields, \begin{equation}
    m_g^2 = \frac{\hbar^2}{c^4} \times \frac{3}{2} H_0^2 \Omega = \frac{\hbar^2}{c^2} \left( \frac{3 H_0^2 \Omega}{c^2} \right)
\end{equation}
which holds for any form of energy density, including contributions from matter or the cosmological constant $\Lambda$. Finally, the cosmological constant $\Lambda$ can be expressed in terms of the density parameter $\Omega_\Lambda$ specific to dark energy, defined as $\Lambda = \frac{3 H_0^2}{c^2} \Omega_\Lambda$. Substituting this definition into the graviton mass relation above, we find that,
\begin{equation} \label{cc2}
    \Lambda = \frac{2 m_g^2 c^2}{\hbar^2}
\end{equation}
We see that \eqref{cc2} is very similar to \eqref{cc} \footnote{One should note that this form of the mass may have some issues with unitarity bounds in proper massive gravity theories \cite{deRham:2010kj,deRham:2014zqa}.}, differing only by a factor of $\frac{4}{3}$.
\\
\\
We can discuss the relation of $m_{g}$ with $\Lambda$ in light of gravitational dispersion relations as well \cite{flauger2018gravitational}. The dispersion relation for general gravitational waves propagating in a medium in this case takes the form\footnote{The reader might be interested to know that there have been other prospective forms of gravitational dispersion relations put forward over the years, for example in \cite{asseo1976general,Garg:2022wdm}.}, \begin{equation} \label{disp2}
    \omega^2 = k^2 c^2 + \frac{32 \pi G \epsilon}{3}
\end{equation}
where $\epsilon$ is the thermal energy density of the medium, which could refer to dark energy or other forms of matter too. For the energy density of dark energy, $\epsilon = \rho_{\Lambda}$, the graviton mass through \eqref{disp2} comes out to be \begin{equation}
    m_{g}^2 = \frac{\hbar^2}{c^4} \frac{32 \pi G \epsilon}{3} 
\end{equation}
Using $\Omega_{\Lambda} = \frac{8 \pi G \rho_{\Lambda}}{3 H_0^2}$ and $\Lambda = \frac{3 H_0^2 \Omega_\Lambda}{c^2} $, we arrive at, \begin{equation} \label{cc3}
    \Lambda = \frac{3 m_{g}^2 c^2}{4 \hbar^2}
\end{equation}
This is similar to \eqref{cc} and \eqref{cc2}, differing by a factor of order unity from both of them. 
\section{Observational Implications}
What are the potential observational signatures here? The Compton wavelength of the graviton,$\lambda_g$ is, \begin{equation}
    \lambda_g = \frac{h}{m_g c} = \frac{2 \pi \hbar}{m_g c}
\end{equation} When considering a relationship between the graviton mass and the cosmological constant in \eqref{cc}, we find that $\lambda_g$ can be further expressed in terms of the Hubble length $l_H = \frac{c}{H_0}$ through the expression,
\begin{equation}
    \lambda_g = \lambda_H \left( \frac{2 \pi^2}{\Omega_\Lambda} \right)^{1/2} = 5.3 l_{h}
\end{equation}
This relationship implies that in the case of \eqref{cc} or \eqref{cc2} or \eqref{cc3} $\lambda_g$ is multiple times larger than $l_H$, the Hubble radius or the observable universe's horizon scale(being around 6 times larger in the case of \eqref{cc} while about 3.5 times for \eqref{cc3} and 5 times larger for \eqref{cc2}). Thus, the Compton wavelength of the graviton significantly exceeds the scale of the observable universe and this has important implications: any physical effects associated with a finite graviton mass in this regime would occur on scales far beyond the reach of observational cosmology, effectively making the graviton mass irrelevant for any cosmological observations or phenomena we could detect within our universe. 
\\
\\
Consequently, any theoretical frameworks that attempt to directly link the cosmological constant to a nonzero graviton mass, either through the linerization approach for \eqref{cc}, the gravitational coupling approach for for \eqref{cc2} or the gravitational dispersion relation approach \eqref{cc3}, faces a fundamental limitation. In other words, a graviton mass in this context would not produce any measurable deviations from general relativity on cosmological scales, rendering such models impractical for explaining observable cosmic acceleration or other large-scale phenomena related to $\Lambda$. This effectively limits the utility of models that try to relate $\Lambda$ to a graviton mass within the scope of observable cosmology, as the predicted graviton mass effects would be undetectable across accessible distances.
\\
\\
To further appreciate the generality of our results here, one may examine extended formulations of the Compton wavelength \footnote{One might note that while these relations were derived in the context of the "reduced" Compton wavelength, the underlying relations of the modified Compton forms hold true. Its essentially a constant rescaling(where $h$ is used instead of $\hbar$ as in \cite{Lake:2015pma,lake2016compton})and the modifications introduced by GUP or higher-dimensional physics are preserved in a scaled-up form that remains mathematically consistent.}. In particular, we explore two modifications of the Compton wavelength: the first is grounded in Generalized Uncertainty Principles (GUP) \cite{Maggiore:1993rv,Adler:2001vs,Tawfik:2015rva,Scardigli:1999jh}, while the second arises from considerations in higher-dimensional physics frameworks. The GUP modifications represent extensions of the Heisenberg uncertainty principle, drawing on theoretical constructs from various quantum gravity paradigms, including string theory, loop quantum gravity, and several facets of black hole physics. The Compton wavelength incorporating GUP with a minimal length scale can be expressed as \cite{Lake:2015pma},
\begin{equation} \label{gup1}
    \lambda = \frac{h}{mc} \left(1 + \alpha \frac{m^2}{m_{p}^2}\right)
\end{equation}
where $m_{p}$ denotes the Planck mass. A further modification for GUP that includes both minimal length and maximal momentum constraints is given by,
\begin{equation} \label{gup2}
    \lambda = \frac{h}{mc \left( 1 - \beta m^2 c^2 \right)}.
\end{equation}
Note that the parameters $\alpha$ and $\beta$ can be either positive or negative, while their estimated values not going much beyond $\mathcal{O}(1)$ \footnote{The interested reader is directed to the following works for varied discussions on these estimates: \cite{amati1989can,maggiore1994quantum,Scardigli:2016pjs,Kanazawa:2019llj,Buoninfante:2019fwr,pedram2012higher,pedram2012higher2}.}. Additionally, one may formulate the Compton wavelength in a (3+n)-dimensional space as \cite{lake2016compton},
\begin{equation} \label{high}
    \lambda = \left( \frac{h}{m c} \right)^{\frac{1}{1+n}} \left( R_{e} \right)^{\frac{n}{1+n}},
\end{equation}
where $R_{e}$ represents a single length scale associated with the compactification of all additional dimensions in which matter may freely propagate, which has the following form for different values of n \cite{arkani1998hierarchy},
\begin{equation}
R_{e} \sim 10^{(32/n) - 17} \text{cm}
\end{equation}
We are now interested in seeing whether the $m_{g}-\Lambda$ relations \eqref{cc},\eqref{cc2} and \eqref{cc3}, could have observationally relevant Compton wavelength estimates after the considerations of these exotic effects. If we take \eqref{cc} for illustration here(the essential results carry over to the other relations too), the GUP modified graviton Compton wavelength \eqref{gup1} takes the form, \begin{equation} \label{l1}
 \lambda_{g} = \sqrt{\frac{2 \pi^2}{\Omega_{\Lambda}}} l_{h} \left( 1 + \frac{2 \alpha G \hbar \Omega_{\Lambda} l_{h}^{-2}}{c^3} \right)   
\end{equation}  
where we used the definition of the Planck mass $m_{p}= \sqrt{\frac{\hbar c}{G}}$. The second term in the brackets on the right hand side in \eqref{l1} is very small, in fact it is so small one can ignore it unless $\alpha$ takes some unrealistically large values. For the case of \eqref{gup2}, the graviton Compton wavelength takes the form, \begin{equation} \label{l2}
    \lambda_{g} = \sqrt{\frac{2 \pi^2}{\Omega_{\Lambda}}} l_{h} \left( \frac{1}{1- 2 \hbar^2 \beta \Omega_{\Lambda} l_{h}^{-2}} \right) 
\end{equation}      
One sees that the Compton wavelength will remain unchanged, as the correction term in the denominator in \eqref{l2} is really small again and can be ignored. This goes to show that even the GUP based corrections to the Compton wavelength, \eqref{gup1}-\eqref{gup2}, are unable to provide observationally relevant results for the graviton. Finally, in the event of higher dimensional corrections to the Compton wavelength \eqref{high} we arrive at the following form, \begin{equation} \label{l3}
    \lambda_g = \left( \sqrt{\frac{2 \pi^2}{\Omega_{\Lambda}}} l_{h} \right)^{\frac{1}{1+n}} R_{e}^{\frac{n}{1+n}}
\end{equation}
It might be tempting to see that the form of n in the exponent could apparently make it possible to reduce the Compton wavelength, which is indeed the case. For example, in case of one extra spatial dimension $n=1$, the graviton Compton wavelength \eqref{l3} $\lambda_g \sim \mathcal{O}(10^{22}) \text{cm}$. This is indeed not superhorizon, however, this falls almost 5 orders of magnitude lesser than the latest constraints on the graviton Compton wavelength inferred from the CMB dipole \cite{loeb2024new}. For higher number of extra dimensions, including the interesting $n=7$ M-theory scale, the Compton wavelength is well below the observational requirements. From this it is clear that the $m_{g}-\Lambda$ correspondence remains observationally irrelevant even after exotic high energy physics considerations.
\acknowledgments
This work was supported in part by Harvard’s Black Hole Initiative, which is funded by grants from JFT and GBMF. The authors would like to thank Daniel Grumiller for some useful comments on the work.

% The bibliography will probably be heavily edited he work.during typesetting.

\bibliography{JSPJMJcitations.bib}

\bibliographystyle{unsrt}

\end{document}